\documentclass[a4paper,fleqn,usenatbib]{mnras}

\usepackage{mathptmx}

\usepackage[T1]{fontenc}
\usepackage{ae,aecompl}

\usepackage{graphicx}	
\usepackage{amsmath}	
\usepackage{amssymb}	

\newcommand{\NII}{[N{\sc ii}]6584~\AA}

\newcommand \vhel{\ifmmode{~V_{{\rm HEL}}}\else{~$V_{{\rm HEL}}$}\fi}
\newcommand \kms{km s$^{-1}$}

\title[On the possible triple central star of PN SuWt~2]{On the possible triple central star system of PN SuWt~2: No m\'enage \`a trois at the heart of the Wedding Ring}
\author[Jones \& Boffin]{David Jones$^{1,2}$\thanks{E-mail:
djones@iac.es}  \& Henri~M.~J. Boffin$^3$ 
\\
$^{1}$Instituto de Astrof\'isica de Canarias, E-38205 La Laguna, Tenerife, Spain\\
$^{2}$Departamento de Astrof\'isica, Universidad de La Laguna, E-38206 La Laguna, Tenerife, Spain\\
$^{3}$European Southern Observatory, Karl Schwarzschild Strasse 2, 85748 Garching, Germany\\
}

\date{Accepted xxxx xxxxxxxx xx. Received xxxx xxxxxxxx xx; in original form xxxx xxxxxxxx xx}

\pagerange{\pageref{firstpage}--\pageref{lastpage}} \pubyear{2016}

\begin{document}
\label{firstpage}
\pagerange{\pageref{firstpage}--\pageref{lastpage}}
\maketitle

\begin{abstract}
SuWt~2 is a planetary nebula consisting of a bright ring-like waist from which protrude faint extended lobes - a morphology believed to be typical of progenitors which have experienced a close-binary evolution.  Previous observations of NSV~19992, the star at the projected centre of SuWt~2, have found it to comprise two A-type stars in a 4.9 day eclipsing orbit, neither of which could be the nebular progenitor.  Radial velocity studies provided a hint that the systemic velocity of this double A-type binary might be varying over time, suggesting the presence of a third component hypothesised to be the nebular progenitor.  Here, we present an extensive radial velocity monitoring study of NSV~19992, performed with the high-resolution echelle spectrograph UVES mounted on ESO's VLT, in order to constrain the possible variation in the systemic velocity of the A-type binary and its relation to the progenitor of SuWt~2.  The observations, acquired over a period of approximately one year, show no evidence of variability in the systemic velocity of NSV~19992.  Combining these new observations with previous high-resolution spectroscopy demonstrates that the systemic velocity is also stable over much longer periods and, moreover, is distinct from that of SuWt~2, strongly indicating that the two are not associated.  We conclude that NSV~19992 is merely a field star system, by chance lying in the same line of sight as the nebular centre, and that it bares no relation to SuWt~2 or its, as yet unidentified, central star(s).
\end{abstract}


\section{Introduction}
\label{sec:intro}

Close-binary interactions are now considered the most-likely mechanism for the formation of aspherical planetary nebulae (PNe), and in PNe where a close-binary nucleus has been detected their shaping influence is clear \citep{hillwig16}.  However, doubts still remain as to whether central star binarity can be responsible for \emph{all} observed aspherical PNe, or whether they are responsible for only a fraction of the total population (perhaps those showing the greatest deviation from sphericity). Conservative estimates place the surviving close binary fraction amongst PN central stars at $\sim$ 20\% \citep{bond00,miszalski09a}, while the true fraction of PNe that will have experienced some form of binary interaction (including mergers) is almost certainly much higher \citep{demarco04,demarco15}.  However, given that approximately 25\% of all solar-type stars occur in triple or higher order systems \citep{raghavan10}, it is possible that such configurations may also be responsible for an appreciable fraction of the PN population \citep{soker94,soker16}.  Based on morphological considerations, \citet{bear16} reason that the fraction of PNe originating from triple systems may be as high as 20\% - comparable to the fraction of detectable close binary central stars.

\begin{figure*}
\centering
\includegraphics[width=\textwidth]{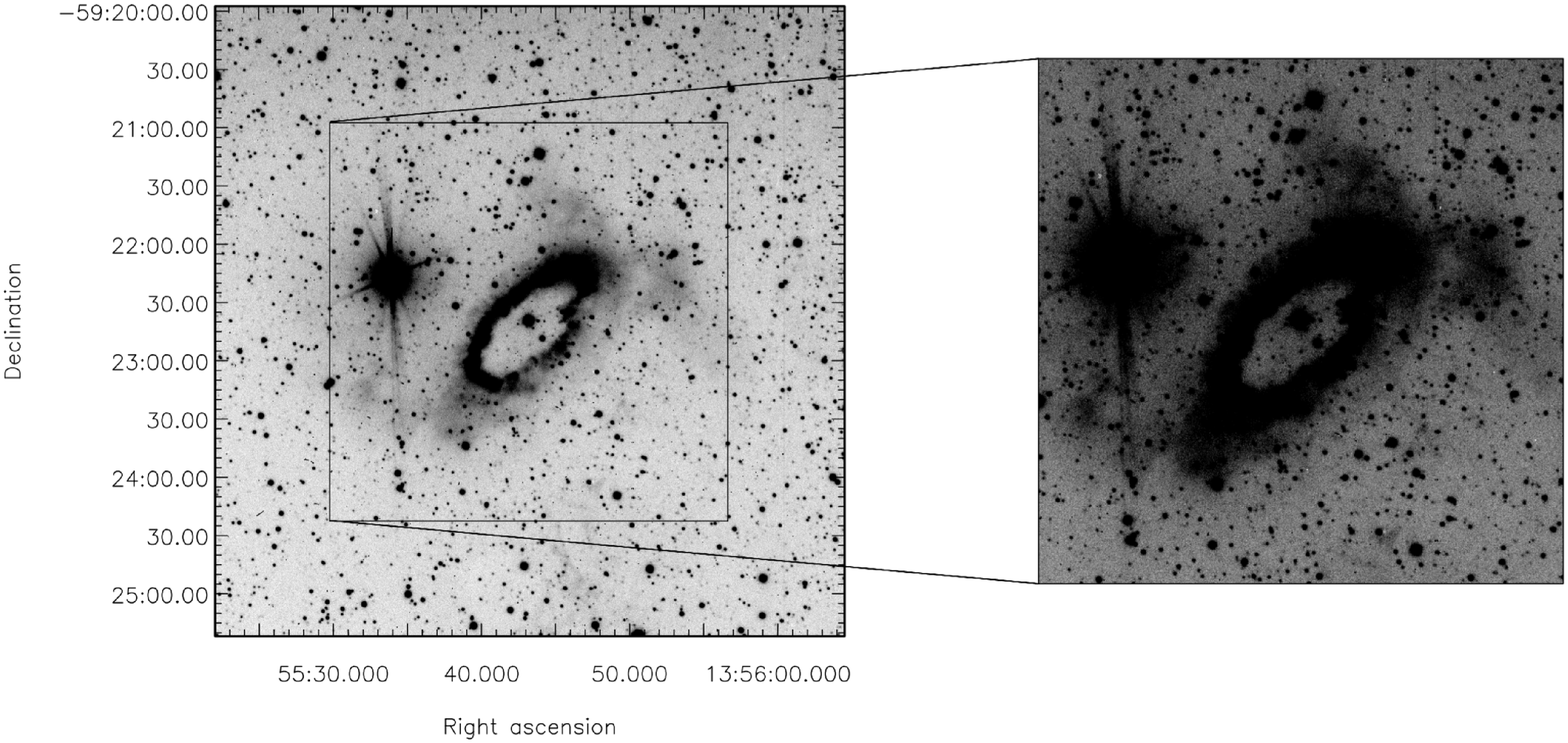}
\caption{ESO-NTT image of SuWt~2 in the light of \NII{} reproduced in part from \citet{jones10a}.}
\label{fig:image}
\end{figure*}

To date, only one PN has been demonstrated to host a triple (or higher-order multiple) star system at its centre, NGC~246 \citep{adam14}. Here, the orbital separation of the nebular progenitor and its closest companion was wide enough to avoid the common-envelope phase, thus limiting the influence of the triple evolution on the formation of the resulting PN.  
The central star system of LoTr~5 was concluded to be a hierarchical triple by \cite{jasniewicz87} and \cite{malasan91}. However, a more extensive, data-rich analysis by \cite{vanwinckel14} found no evidence of radial velocity variations associated with three orbital components, instead finding that the radial velocity variability is due to a wide-binary orbit with a period of a few thousand days.  The central star of M2-29 has also been suggested to be a hierarchical triple where the nebular progenitor is in a 23-day period orbit with a close companion and that they in turn orbit a tertiary component with an 18 year period \citep{hajduk08}.  \cite{miszalski11e}, however, find no evidence for a tertiary component, demonstrating that the observed periodic changes in colour are associated purely with dust obscuration events rather than a wide companion.

The strongest remaining candidate is, without doubt, PN SuWt~2 \citep{schuster76}, a ring nebula (see figure \ref{fig:image}) where the bright star closest to the projected nebular centre has been shown to be a 4.9 day period binary consisting of two, almost identical, A-type main-sequence stars \citep{bond02,exter03}.  \citet{exter10} performed a detailed analysis of this double A-type system \citep[NSV~19992 from hereon; ][]{kazarovets98}, showing both components to be A1V stars of mass  $\sim$2.7M$_\odot$, neither of which could be the source of photoionisation nor the nebular progenitor \citep[the nebular abundances indicate a post-AGB progenitor; ][]{smith07}.  Furthermore, their rotational velocities were found to be different and much slower than expected if the two were tidally locked into rotating synchronously with the orbital period.  \citet{exter10} argue that this may result from the previous interaction with a currently-unseen post-AGB companion, which is both the nebular progenitor and source of photoionisation, but which was below the detection limit of their UV observations.  The most convincing evidence for association between NSV~19992 and the unseen progenitor of SuWt~2 is in the form of apparent variability in the systemic velocity of NSV~19992. \citet{exter10} combined spectroscopy from multiple telescopes and instruments taken over a period of nearly 6 years, finding heliocentric systemic velocities ranging from approximately $-43$ to $-4$\kms{} \citep[an amplitude of $\approx$20 \kms{} around the systemic velocity of the nebula as determined by][]{jones10a}.  They note that (assuming an orbital period of $\approx$0.2 years, implied by the shortest separation between their radial velocity measurements) the putative third-body in such a system would have an unfeasibly large mass (greater than the Chandrasekhar mass).  However, by discarding an outlying measurement (based on low-resolution data), they could revise the radial velocity amplitude down to $\approx$10\kms{}, resulting in an estimated mass for the tertiary component of 0.7M$_\odot$.  \citet{danehkar13} derived a photoionisation model for the nebula and its central star, finding that a central star of mass $\sim$0.7M$_\odot$ could result in the measured nebular line fluxes for a distance (and implied nebular size and post-AGB age) consistent with that determined by \cite{exter10} for NSV~19992.

The parallax of NSV~19992 determined by \textit{Gaia} is 0.655 $\pm$ 0.277 mas \citep{gaia_satellite}, equivalent to a distance of 0.3--2.7 kpc (5--95\% confidence interval), consistent with the distance determination of \citet{exter10} at $D=2.3\pm0.2$ kpc (found by modelling of the binary).  The only distance determination for the PN SuWt~2 based on its nebular properties alone (and not the properties of NSV~19992) is that of \citet{frew16} using their H$\alpha$ surface brightness - radius relation, where an optically thick distance of 3.78$\pm$1.11 kpc was found\footnote{\citet{frew16} determine a statistical ``mean'' distance of 3.18$\pm$0.93 kpc for SuWt~2 (more or less consistent with the distance of NSV~19992), but state that, given the properties of the nebula, the optically thick value of 3.78$\pm$1.11 kpc should be used.}, only marginally consistent with the distance of the A-type binary NSV~19992 (in spite of the fact that SuWt~2 is used as a calibrator in their determination by assuming that both NSV~19992 is, in fact, related to SuWt~2 and that their common distance is that determined by \citealt{exter10}).  This may be a reflection of the inherent uncertainties in statistical distance determinations for PNe or an indication that NSV~19992 and SuWt~2 are not related.  Furthermore, it is interesting to note that NSV~19992 is not at the projected, geometric centre of SuWt~2 \citep[offset by $\sim$1\arcsec{};][]{smith07,jones10a}, which may be considered evidence that, indeed, the two are unrelated. However, many PN central stars are seen displaced from the geometric centres \citep[e.g. the central star of Abell~41, MT Ser; ][]{jones10b} and this may be a possible consequence of interactions between the central star components \citep{soker98a}.

Here, we present a detailed radial velocity analysis of NSV~19992 based on high-resolution spectra taken with the same instrumental set-up (to avoid possible systematics) over a period of almost one year, in order to constrain the possible variability in the systemic velocity of the binary and assess its relationship with PN SuWt~2.  The observations and analysis are presented in section \ref{sec:obs}, while the conclusions are presented in section \ref{sec:conclusions}.

\section{Observations and analysis}
\label{sec:obs}

\begin{table}
\caption{A table of measured heliocentric radial velocities versus heliocentric modified Julian date for the two components NSV~19992.}
\label{tab:rvs}
\begin{tabular}{lrlrl}
\hline
 Modified Julian & \multicolumn{2}{c}{Radial velocity of} & \multicolumn{2}{c}{Radial velocity of }\\
  date (days) & \multicolumn{2}{c}{primary (\kms{})} & \multicolumn{2}{c}{secondary (\kms{})}\\
\hline
56737.349454 	 & 	 $-$91.118 	 & 	 $\pm$0.689 	 & 	 75.843 	 & 	 $\pm$0.272\\
56762.160084 	 & 	 $-$110.837 	 & 	 $\pm$0.624 	 & 	 94.590 	 & 	 $\pm$0.272\\
56773.060922 	 & 	 $-$61.900 	 & 	 $\pm$0.651 	 & 	 47.622 	 & 	 $\pm$0.296\\
56776.139399 	 & 	 $-$36.705 	 & 	 $\pm$0.736 	 & 	 19.060 	 & 	 $\pm$0.291\\
56779.009193 	 & 	 68.177 	 & 	 $\pm$0.677 	 & 	 $-$85.180 	 & 	 $\pm$0.288\\
56784.999457 	 & 	 80.965 	 & 	 $\pm$0.760 	 & 	 $-$99.180 	 & 	 $\pm$0.257\\
56787.004932 	 & 	 $-$115.839 	 & 	 $\pm$0.583 	 & 	 100.550 	 & 	 $\pm$0.257\\
56789.232732 	 & 	 96.686 	 & 	 $\pm$0.558 	 & 	 $-$113.426 	 & 	 $\pm$0.260\\
56792.084737 	 & 	 $-$113.517 	 & 	 $\pm$0.574 	 & 	 97.896 	 & 	 $\pm$0.267\\
56811.025878 	 & 	 $-$92.917 	 & 	 $\pm$0.778 	 & 	 78.505 	 & 	 $\pm$0.298\\
56813.987319 	 & 	 99.210 	 & 	 $\pm$0.636 	 & 	 $-$117.322 	 & 	 $\pm$0.287\\
56831.038256 	 & 	 $-$113.780 	 & 	 $\pm$0.704 	 & 	 99.735 	 & 	 $\pm$0.313\\
56833.993277 	 & 	 90.789 	 & 	 $\pm$0.866 	 & 	 $-$105.689 	 & 	 $\pm$0.336\\
56838.986874 	 & 	 83.747 	 & 	 $\pm$0.948 	 & 	 $-$100.173 	 & 	 $\pm$0.333\\
56841.000608 	 & 	 $-$114.474 	 & 	 $\pm$0.743 	 & 	 101.932 	 & 	 $\pm$0.329\\
56848.117489 	 & 	 95.477 	 & 	 $\pm$0.697 	 & 	 $-$112.079 	 & 	 $\pm$0.322\\
56870.127909 	 & 	 $-$107.291 	 & 	 $\pm$0.896 	 & 	 92.801 	 & 	 $\pm$0.365\\
56874.037974 	 & 	 10.415 	 & 	 $\pm$1.507 	 & 	 $-$21.500 	 & 	 $\pm$0.335\\
56876.069376 	 & 	 $-$74.394 	 & 	 $\pm$0.835 	 & 	 60.670 	 & 	 $\pm$0.347\\
56881.999839 	 & 	 55.841 	 & 	 $\pm$0.823 	 & 	 $-$71.640 	 & 	 $\pm$0.314\\
56885.033362 	 & 	 $-$112.847 	 & 	 $\pm$0.745 	 & 	 100.015 	 & 	 $\pm$0.338\\
56887.018940 	 & 	 68.068 	 & 	 $\pm$0.720 	 & 	 $-$83.394 	 & 	 $\pm$0.311\\
56897.039391 	 & 	 86.259 	 & 	 $\pm$0.928 	 & 	 $-$100.739 	 & 	 $\pm$0.354\\
56903.044846 	 & 	 63.119 	 & 	 $\pm$0.766 	 & 	 $-$78.646 	 & 	 $\pm$0.293\\
57057.366174 	 & 	 $-$105.946 	 & 	 $\pm$0.665 	 & 	 93.962 	 & 	 $\pm$0.280\\
57087.171205 	 & 	 $-$76.401 	 & 	 $\pm$0.747 	 & 	 64.590 	 & 	 $\pm$0.323\\
\hline
\end{tabular}
\end{table}

NSV~19992 was observed 26 times with the Ultraviolet and Visual Echelle Spectrograph \cite[UVES; ][]{dekker00} on the Kueyen Unit Telescope of the European Southern Observatory's Very Large Telescope (ESO-VLT) at part of the service mode programme 593.D-0037 between March 21 2014 and March 6 2015 (see table \ref{tab:rvs} for the exact dates of observation).  The instrument set-up was the same for each observation, employing a 0.6\arcsec{} slit and the dichroic \#1.  Simultaneous 1200-s exposures were taken in both arms, with the blue arm centred at 3900\AA{} (spectral range $\sim$3400--4500\AA{}, with the standard HER\_5 below-slit filter) and the red at 5800\AA{} (spectral range $\sim$4800--6800\AA{}, with the standard SHP700 below-slit filter).  The resulting data were then processed with the standard ESO pipeline and downloaded from the ESO Science Archive Facility.

\begin{figure}
\centering
\includegraphics[width=\columnwidth]{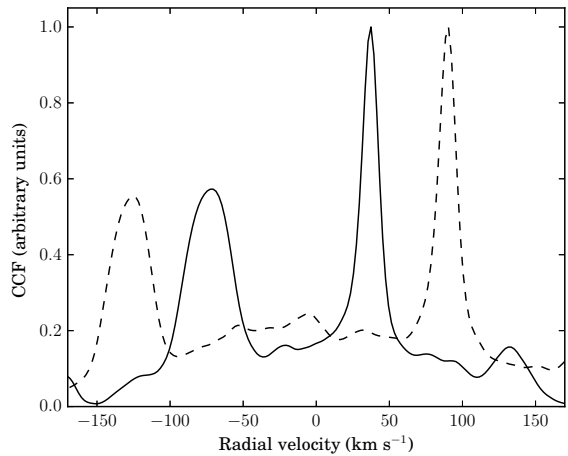}
\caption{Two example CCFs, corresponding to observations taken at MJD 56773.061 (solid line) and 56787.005 (dashed line), clearly displaying the obvious peaks associated with the two components of NSV~19992.  The different peak widths reflect the different rotational velocities of the two components.}
\label{fig:ccf}
\end{figure}

The spectra were then continuum subtracted, corrected to heliocentric velocity and cross-correlated against a synthetic A1V spectrum \citep[with the same parameters as those derived by][namely T$_{eff}$ = 9250 K and log $g$ = 4]{exter10} produced using the \textsc{spectrum} software \citep{spectrum}.
Each cross-correlation function (CCF) displayed a clear, double-peaked profile, thus allowing us to easily discern the radial velocities of both components of the binary by gaussian fitting of the individual CCFs (example CCFs are shown in figure \ref{fig:ccf}). 
The resulting heliocentric radial velocities and their associated uncertainties are listed in table \ref{tab:rvs}.

\begin{figure*}
\centering
\includegraphics[width=\textwidth]{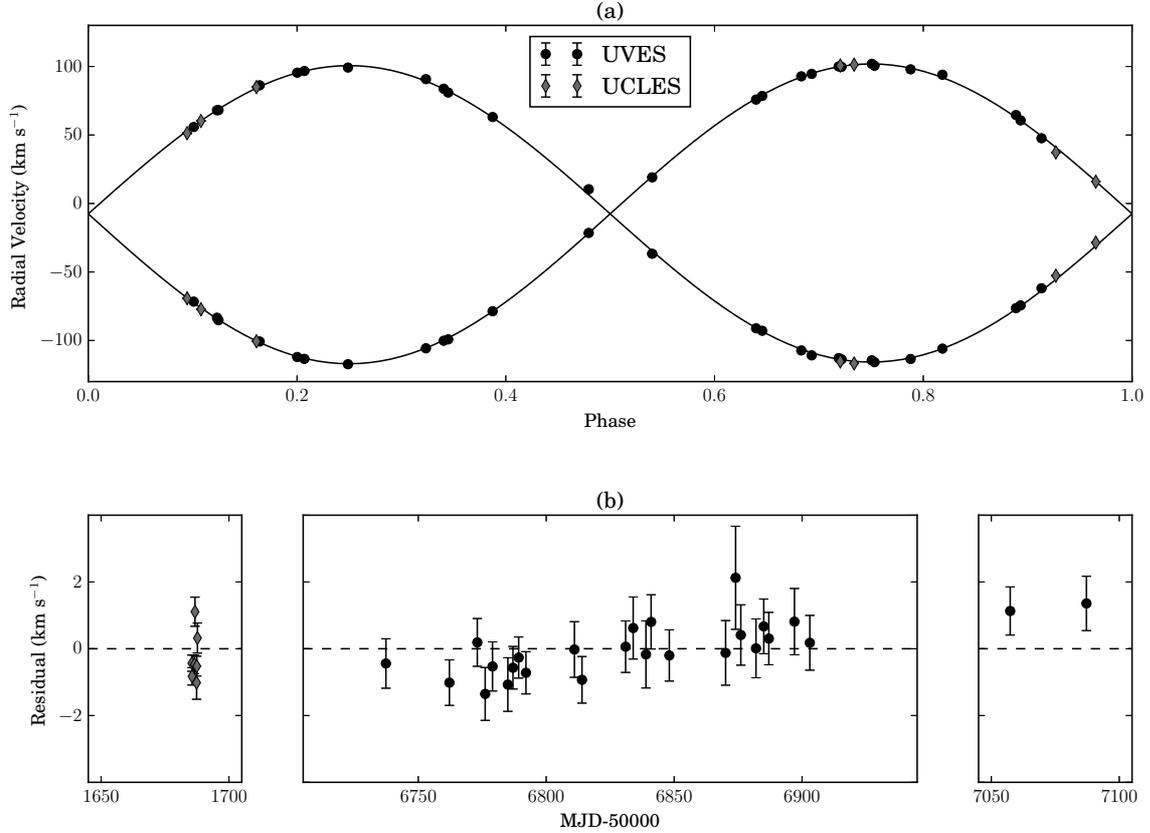}
\caption{Upper panel: Folded RV curve of NSV~19992 showing a good fit with no appreciable shift in $\gamma$ between epochs \citep[circular filled points are the new UVES data points, while the diamonds are data from AAT-UCLES reproduced from][]{exter10}. Lower panel: The residuals for each epoch of UVES and UCLES data (c.f.\ the sine fit shown in the upper panel).
}
\label{fig:RVcurve}
\end{figure*}

\begin{table}
   \centering
   \caption{Orbital parameters}
   \label{tab:params}
   \begin{tabular}{lc}
      \hline\hline
Parameter & Value\\
\hline
$q$ & 0.990 $\pm$ 0.007\\
$K_1$ (\kms{}) & 108.23 $\pm$ 0.39\\
$K_2$ (\kms{}) & 109.28 $\pm$ 0.31\\
$\gamma$ (\kms{}) & $-$7.60 $\pm$ 0.21\\
$P$ (days) & 4.9098505*\\
$T_0$ (HJD) & 2456842.7248 $\pm$ 0.0010\\
      \hline
   \end{tabular}
   \medskip
  
   *Fixed parameter taken from \cite{exter10}
\end{table}

The radial velocities of the two components of NSV~19992 were simultaneously fit in order to check for signs of variability in the systemic velocity, $\gamma$, of the binary.  The period was fixed to the value determined by \citet{exter10}, who used several epochs of photometric data taken over approximately ten years in order to derive an extremely precise period (4.9098505$\pm$0.0000020 days).  The data were satisfactorily fit with a solution containing no drift in systemic radial velocity, with residuals generally of the order of the uncertainty on the radial velocity measurement at each epoch ($\sim$1 \kms{}).  The resulting folded radial velocity curve and associated fit is shown in figure \ref{fig:RVcurve}(a), while the residuals to this fit as a function of MJD are shown in figure \ref{fig:RVcurve}(b).  The parameters of the fit, and the associated derived quantities for NSV~19992 are presented in table \ref{tab:params}.

It is immediately clear by looking at the residuals to the fit that our results do not replicate the large, $\sim$20 \kms{} over a period of $\sim$0.2 years, amplitude variations in systemic velocity found by \citet{exter10}.  Our data show no signs of any deviation greater than $\sim$2 \kms{} over the full observing period ($\sim$1 year) implying that the period of any third body (with reasonable mass for a post-AGB stellar remnant) would be of order several years or more (see figure \ref{fig:Periods}).  The residuals (figure \ref{fig:RVcurve}(b)) do appear to show a general upward trend throughout the observing period, however once the uncertainties on the sinusoidal fits are taken into account this trend is not statistically significant (sinusoidal fitting of the residuals results in solutions where the derived amplitudes are less than or approximately equal to the uncertainty of the fit, i.e.\ a less than 1$\sigma$ correlation).  The fit is found to be equally good to the new data presented here as for the AAT-UCLES data of \citet{exter10}, with residuals\footnote{Note that the amplitude of the scatter of the UCLES points is more or less the same as the ``drift'' among the UVES points, further evidence that this apparent, low-level variability is just $\sim$1$\sigma$ deviations about the mean.}  generally $\lesssim$1 \kms{} as shown in the left-hand panel of figure \ref{fig:RVcurve}(b).  The agreement is further demonstrated when comparing the derived systemic velocities from our fit to the UVES data ($-$7.60$\pm$0.21 \kms{}) and that derived by \citet{exter10} when solely considering their AAT-UCLES data ($-$7.82$\pm$0.41 \kms{}).
It would, indeed, be an extreme coincidence that the only epochs at which the systemic velocity of NSV~19992 has been measured with high-resolution, cross-dispersed echelle data reflect the same phase of the third body's orbit, as such we are led to conclude that the previous radial velocity measurements based on gaussian fitting of single absorption lines in medium-resolution spectra do not accurately reflect the true radial velocity of the system at those epochs.  Furthermore, the fact that the residuals to our data (covering almost one year, with good sampling throughout) are of the same order as to the AAT-UCLES data taken some 13 years earlier, is strongly indicative that the systemic velocity of NSV~19992 is constant on both short and long timescales.  Similarly, the constant systemic velocity for both data sets makes a tertiary component in an eccentric orbit rather unlikely.  Attempts were made to fit to an eccentric orbit including the data points presented here as well as all data from \citet{exter10} in order to evaluate the possibility of a long-period with high eccentricity, however no satisfactory solution could be found.  Even selectively excluding data points from lower resolution instruments (for example, just using UVES, UCLES and SAAO data), the long baseline of the (non-variable) UVES data seemingly excludes any eccentric solutions.

In addition to their AAT-UCLES data, the AAT-RGO data from \citet{exter10} are also found to agree relatively well with the derived fit (with residuals again of the order of the uncertainties).  However, it is important to note that the fit presented here does not reproduce the radial velocities measured by \citet{exter10} from their lower resolution SAAO and NTT data. This is perhaps not unexpected given that inclusion of all data (high-, medium- and low-resolution) provided \citet{exter10} with a minimum radial velocity semi-amplitude that implied an unfeasibly high central star mass (greater than the Chandrasekhar mass), which led them to ignore the lowest resolution data in estimating their proposed orbital solution.  The orbital semi-amplitudes, $K_1$ and $K_2$, for our fit are systematically lower than those found by \cite{exter10} -- this can be explained by the different  $T_0$ we obtain (although our values agree within the uncertainties), resulting in a slight difference in phase for the resulting radial velocity fit.  However, the measured mass ratio, $q$, is found to be very close to unity in good agreement with the value determined by \cite{exter10}.  We find no evidence for any deviation from a circular orbit, just as concluded by \cite{exter10} from their extensive eclipse timings (while their radial velocity fit did seem to indicate a statistically significant non-zero eccentricity).

\section{Conclusions}
\label{sec:conclusions}

\begin{figure}
\centering
\includegraphics[width=\columnwidth]{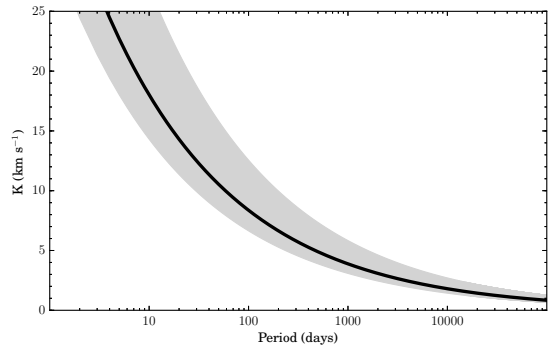}
\caption{Period versus maximum observed radial velocity ($K$) for a central star mass, $M_{\mathrm{CS}}$=0.64 M$_\odot$, as derived by \citet{danehkar13} and assuming an inclination of 68\degr{} \citep{jones10a} and the most conservative case of a circular orbit.  The shaded region encompasses the range 0.5 $<$ M$_{\mathrm{CS}} <$ 1.0 M$_\odot$ to demonstrate the relatively weak dependence on central star mass. 
}
\label{fig:Periods}
\end{figure}

The systematic velocity of NSV~19992, based on our data and the echelle data of \citet{exter10}, is not found to vary with time.  The residuals to a sinusoidal fit are of order $\lesssim$1 \kms{}, placing very strong limits on the amplitude of radial velocity variability due to any possible third component to the system.  Assuming a reasonable mass for a post-AGB stellar remnant capable of producing and then ionising SuWt~2, any radial velocity variability hidden in the fit residuals would imply an orbital period of several tens of years (see figure \ref{fig:Periods}), certainly not consistent with any kind of common-envelope evolution involving NSV~19992 (therefore greatly limiting the shaping influence of any possible triple configuration).  Furthermore, the systemic velocity of NSV~19992, and therefore of any triple system comprised of NSV~19992 and the nebular progenitor (given that large radial velocity variations in NSV~19992 can be ruled out), is significantly different from the systemic velocity of PN SuWt~2 \citep[$-7.6$ c.f.\ $-25$ \kms{}; ][]{jones10a}.  This is strongly indicative that the NSV~19992 and SuWt~2 are not associated as there is little reason to expect that the progenitor star system and resulting PN would have different systemic velocities. The possible discrepancy in distances between NSV~19992 and SuWt~2 highlighted in section \ref{sec:intro} adds further weight to the argument that the two are not connected, as does the fact that the morphology of the nebula would be considered more typical of a binary star evolution \citep{bear16}.  As such, we conclude that NSV~19992 is merely a field system and bears no relation to SuWt~2 or its, as yet, unidentified progenitor.

\section*{Acknowledgments}
Based on observations made with ESO Telescopes at the La Silla Paranal Observatory under programme IDs 55.D-0550, 74.D-0373, 593.D-0037. This research has made use of NASA's Astrophysics Data System Bibliographic Services; the SIMBAD database, operated at CDS, Strasbourg, France; the VizieR catalogue access tool, CDS, Strasbourg, France; APLpy, an open-source plotting package for Python hosted at http://aplpy.github.com; Astropy, a community-developed core Python package for Astronomy \citep{astropyshort}; PyAstronomy; SciPy \citep{scipy}; NumPy \citep{numpy}.  This work has made use of data from the European Space Agency (ESA) mission {\it Gaia} (\url{http://www.cosmos.esa.int/gaia}), processed by the {\it Gaia} Data Processing and Analysis Consortium (DPAC, \url{http://www.cosmos.esa.int/web/gaia/dpac/consortium}). Funding for the DPAC has been provided by national institutions, in particular the institutions participating in the {\it Gaia} Multilateral Agreement.

\bibliographystyle{mn2e}
\bibliography{literature.bib}

\label{lastpage}

\end{document}